\documentclass[final,1p,times,twocolumn]{elsarticle}

\usepackage{amssymb}

 \usepackage{lineno}

\usepackage{lineno,hyperref}
\modulolinenumbers[5]

\journal{Asociaci\'on Argentina de Astronom\'ia
BAAA, Vol. 54, 2011}

\begin{document}

\begin{frontmatter}

\title{\textbf{Observaci\'on de lentes gravitatorias con ALMA}}

\author[label]{A. Bonilla\footnote{alex.acidjazz@gmail.com}}
\address[label]{Departamento de F\'isica y Astronom\'ia, Facultad de Ciencias, Universidad de Valpara\'iso, Valpara\'iso, Chile.
\\}

\author[label]{O. Toloza}
\author[label]{I. Fuentes}
\author[label]{V. Motta}

\begin{abstract}
Gravitational lensing is a fundamental tool for cosmology. A recent instrument which will provide more information for models of these objects is ALMA. Our goal is to select lens candidates to observe with ALMA and then model them using GravLens Software. We had selected 12 quadruple images systems from the CASTLES database, which show a high probability of observ- ing extended sources in the submillimetric range. These new data will allow us to improve existing models.\\

\noindent \textbf{Resumen}\\

\noindent  Las lentes gravitatorias son una herramienta fundamental para la cosmolog\'ia. Un nuevo instrumento que nos proporcionar\'a mayor informaci\'on para los modelos de estos objetos, es ALMA. Nuestro objetivo es seleccionar lentes candidatas para observar con ALMA y posteriormente modelarlas mediante el programa \textit{GravLens}. Seleccionamos de la base de datos de CASTLES 12 sistemas cu\'adruples, los cuales tienen mayor probabilidad de observar fuentes extendidas en el rango submilim\'etrico. Estos nuevos datos nos permitir\'an mejorar los modelos exitentes para dichos sistemas.
\end{abstract}

\begin{keyword}
\texttt{ Lentes gravitatorias, ALMA, submilim\'etrico.}
\MSC[2011] 00-01\sep  99-00
\end{keyword}

\end{frontmatter}


\section{Introduction}\label{sec:01}

\noindent 
El efecto lente gravitatoria se produce cuando la luz de un objeto distante (fuente) se desv\'ia, debido a una distribuci\'on de masa (lente), produciendo im\'agenes de la fuente  \cite{Refsdal:02}. Uno de los instrumentos que nos proporcionar\'a mayor informaci\'on sobre estos sistemas es el \textit{Atacama Large Millimeter/submillimeter Array} (ALMA) \cite{ALMA:05}. Actualmente, en el ciclo cero se encuentran operativas 12 antenas para las configuraciones compacta y extendida con receptores cuyas bandas disponibles son las denominadas 3, 6, 7 y 9, que cubren un rango de frecuencia de 84-720 GHz (MANUAL) \cite{MANUAL:06}.

\begin{figure}[htb]
 \centering
 \begin{center}
\includegraphics[height=4cm]{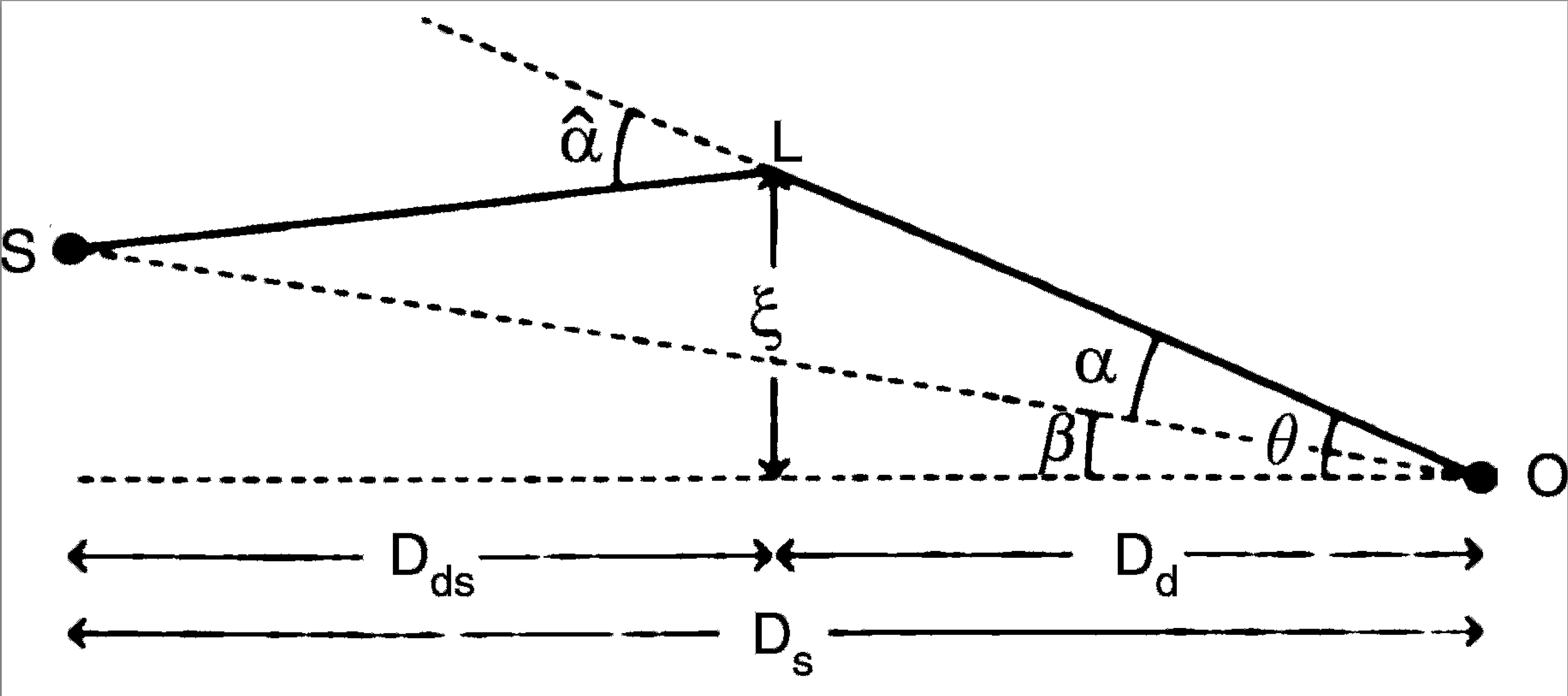} 
\caption{Aproximaci\'on geom\'etrica del sistema lente para \'angulos peque\~nos. $D_{d}$, $D_{s}$ y $D_{ds}$ son la distancias a la lente, a la fuente, y la distancia lente-fuente, respectivamente.}\label{fig1}
\end{center}
\end{figure}

\section{Marco te\'orico del efecto lente}\label{sec:02}

\noindent
Utilizando el esquema de la \textbf{figura \ref{fig1}} puede deducirse la ecuaci\'on de la lente:

\begin{equation}
\beta=\theta - \alpha(\theta).
\end{equation}

\noindent Esta ecuaci\'on relaciona la posici\'on real de la fuente ($\beta$) con la posici\'on de la imagen ($\theta$) y el \'angulo de deflexi\'on escalado ($\alpha$). Se definen \textit{densidad de masa superficial} y \textit{densidad de masa superficial cr\'itica} como: 

\begin{equation}
\Sigma\equiv \int \rho(r) dr, \quad \Sigma_{crit}=\frac{c^{2}}{4 \pi G} \frac{D_{s}}{D_{d}D_{ds}},
\end{equation}

\noindent donde $D_{s}$, $D_{d}$ y $D_{ds}$ son las distancias a la fuente, a la lente y entre la fuente y la lente respectivamente (ver Schneider, 2006, 122) \cite{Schneider:03}. Para calcular estas distancias usamos el modelo cosmol\'ogico \textit{``Lambda Cold Dark Matter''}, ($\mathrm{\Lambda}$CDM), cuyos par\'ametros son $\mathrm{\Omega_{m}} \approx 0.28$, $\mathrm{\Omega_{\Lambda}} \approx 0.72$, $\mathrm{\Omega_{k}} \approx 0$.\\

\begin{figure}[htb]
 \centering
\includegraphics[height=4.5cm]{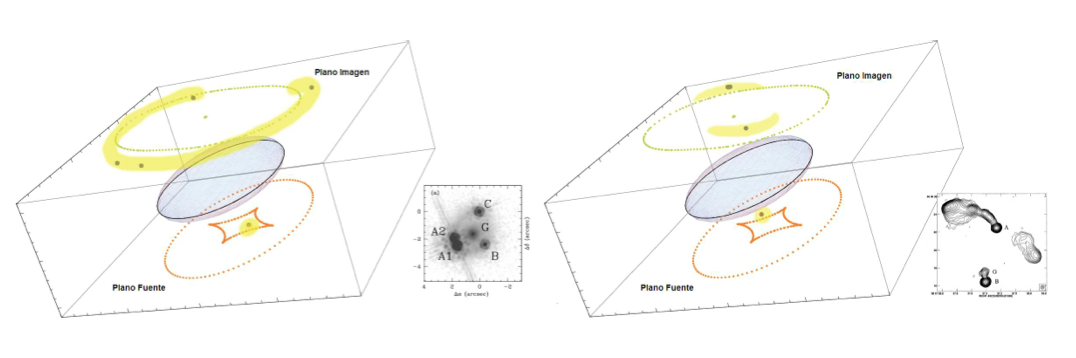} 
\caption{Esquema de las posiciones de las im\'agenes con respecto al plano de la lente (verde) y su correspondiente mapeo en el plano de la fuente (naranja). Las fuentes extendidas y puntuales se muestran en amarillo y azul, respectivamente. \textit{Izquierda}: Sistema de 4 im\'agenes formado cuando la fuente se encuentra dentro de la c\'austica. \textit{Derecha}: Sistema de 2 im\'agenes formado cuando la fuente se encuentra fuera de la 
c\'austica.}\label{fig2}
\end{figure}

\noindent La curva cr\'itica corresponde a la regi\'on donde la magnificaci\'on de las im\'agenes es m\'axima y la curva c\'austica corresponde al mapeo de tal curva cr\'itica por medio de la ecuaci\'on de la lente. La formaci\'on de las im\'agenes depende de la posici\'on de la fuente respecto a la curva c\'austica. Se observan 2 \'o 4 im\'agenes si la fuente se encuentra fuera o dentro de la c\'austica, respectivamente (ver \textbf{figura \ref{fig2}} ). Cuando observamos un sistema cu\'adruple formando arcos es debido a que las regiones externas de una fuente extensa se encuentran sobre la c\'austica. \\

\begin{figure}[htb]
 \centering
\includegraphics[height=4cm]{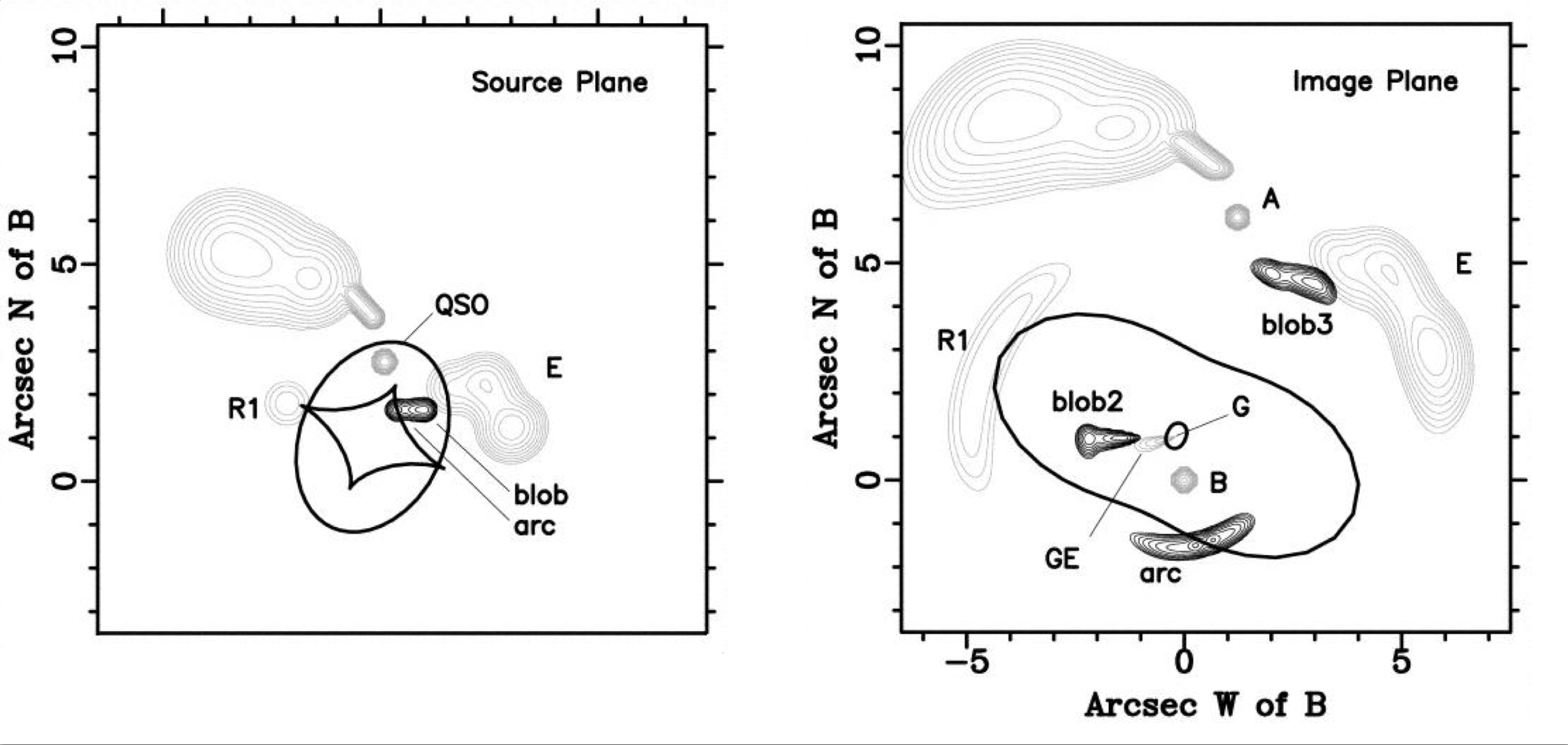} 
\caption{Representaci\'on del plano de la fuente (izquierda) y de la imagen (derecha) de QSO 0957+561. Se muestran las curvas c\'austicas y cr\'iticas en cada plano, y la verdadera posici\'on de la fuente modelado con \textit{GravLens}. Tambi\'en se observa un l\'obulo extendido y un radio jet (Wucknitz, 2007) \cite{Wucknitz:04}.}\label{fig3}
\end{figure}

\noindent Ejemplo de generaci\'on de arcos es el sistema mostrado en la \textbf{figura \ref{fig3}}. Se observa una fuente extendida que cubre parte de la c\'austica (blob) dando origen a dos im\'agenes y un arco, adem\'as un sistema puntual (QSO) situado fuera de la c\'austica el cual genera dos im\'agenes puntuales.

\section{Modelado de sistemas lentes gravitatorias}\label{sec:03}
\noindent Para modelar el sistema usaremos \textit{GravLens Software Package v1.07} \cite{GravLens:07} con un perfil de distribuci\'on de masa Elipsoide Singular Isotermo. Este perfil se describe por la densidad de masa $\rho$ y la dispersi\'on de velocidad $\sigma_{v}$ las cuales est\'an relacionadas por:

\begin{equation}
\rho(r)=\frac{\sigma_{v}}{2 \pi G r^{2}}, \quad \Sigma(\xi)=\frac{\sigma_{v}^{2}}{2 G} \xi^{-1},
\end{equation}

\noindent donde $\Sigma(\xi)$ es la masa proyectada y $\xi$ es el par\'ametro de impacto (\textbf{figura \ref{fig1}}). Los par\'ametros de entrada de \textit{GravLens} (Keeton, 2002) \cite{Keeton:01}
 corresponden: orientaci\'on, posici\'on y elipticidad de la lente, y las posiciones de las  im\'agenes. Como resultado el programa entrega: la distorsi\'on y su orientaci\'on, la   dispersi\'on de velocidad y la posici\'on de la fuente. Un sistema cu\'adruple en el  \'optico, con emisi\'on extensa en radio, generar\'a m\'as datos de entrada para el modelo.  Adem\'as, a menor frecuencia la se\~nal de las observaciones mejorar\'a debido a que los  efectos de extinci\'on del polvo y microlente son despreciables.

\section{Proyecciones con ALMA}\label{sec:04}

\noindent Usando la base de datos de CASTLES \cite{CASTLES:08}, encontramos 54 lentes observables con ALMA. En base al corrimiento al rojo de la fuente estimamos la frecuencia a la cual emitir\'ia el $^{12}$CO para sus diferentes transiciones. Con esta muestra seleccionamos los sistemas con 4 im\'agenes de la fuente en el \'optico (ver \textbf{tabla \ref{tab1}}), ya que estos sistemas presentan una alta probabilidad de formar arcos en el submilim\'etrico.

 \begin{table}[t]
 \centering
 \begin{tabular}{c|l|c|l|l}
 \hline
 \hline
$nº$ lente & Nombre & Banda & $Z_{lente}$ & $Z_{fuente}$\\

 \hline

10	&HE0230-2130  	&6,7  	&2.162  &0.52\\
14  &MG0414+0534  	&6    	&2.64   &0.96 \\
15  &HE0435-1223  	&6,7  	&1.689  &0.46\\
29  &RXJ0911+0551	&6		&2.80	&0.77\\
32	&SDSS0924+0219	&6,7	&1.524	&0.39\\
47	&PG1115+080		&6,7	&1.72	&0.31\\
49	&RXJ1131-1231	&7		&0.658	&0.295\\
51	&SDSS1138+0314	&6		&2.44	&0.45\\
84	&PMNJ1838-3427	&6		&2.51	&0.89\\
89	&MG2016+112		&6		&3.27	&1.01\\
91	&WFI2033-4723	&6,7	&1.66	&0.66\\
98 	&Q2237+030		&6,7	&1.69	&0.04\\

 \hline
 \end{tabular}
 \caption{Lentes candidatas a observar con ALMA.}\label{tab1}
 \end{table}
 
 \noindent De las bandas disponibles en el ciclo 0 es necesario utilizar dos de ellas para la observaci\'on de los sistemas listados en la \textbf{tabla \ref{tab1}}, debido a que la frecuencia de emisi\'on del $^{12}\mathrm{CO}$ para los sistemas coinciden s\'olo con tales bandas. Informaci\'on de ellas se detalla en la \textbf{tabla \ref{tab2}}.

 \begin{table}[!h]
 \centering
 \begin{tabular}{c|c|c|c}
 \hline
 \hline
 Banda & Rango de frecuencia & Resoluci\'on angular & Campo de visi\'on \\
       &    (GHz)            &   ('')               & ('') \\
 \hline
6	&211-275  	&0.68  	&27 \\
7   &275-373  	&0.45  	&18 \\
 \hline
 \end{tabular}
 \caption{Detalle de las bandas 6 y 7 de la configuraci\'on extendida de ALMA. }\label{tab2}
 \end{table}

\section{Conclusiones}\label{sec:05}
 De los sistemas lentes observables con ALMA seleccionamos 12 sistemas c\'uadruples, los cuales presentan una alta probabilidad de que sean fuentes extendidas en el submilim\'etrico, dada la certeza de que se encuentran dentro de la c\'austica. Estas estructuras nos proveer\'ian de m\'as datos para para mejorar los modelos actuales.

\section*{Agradecimientos}\label{sec:06}

A. Bonilla, O. Toloza, I. Fuentes agradecen el financiamiento de la SOCHIAS a trav\'es del proyecto Gemini-CONICYT 32100009 y al proyecto FONDECYT 1090673. V. Motta agradece el financiamiento del proyecto FONDECYT 1090673.


\def \aap {A\&A} 
\def \aapr {A\&AR} 
\def \statisci {Statis. Sci.} 
\def \physrep {Phys. Rep.} 
\def \pre {Phys.\ Rev.\ E.} 
\def \sjos {Scand. J. Statis.} 
\def \jrssb {J. Roy. Statist. Soc. B} 
\def \pan {Phys. Atom. Nucl.} 
\def \epja {Eur. Phys. J. A} 
\def \epjc {Eur. Phys. J. C} 
\def \jcap {J. Cosmology Astropart. Phys.} 
\def \ijmpd {Int.\ J.\ Mod.\ Phys.\ D} 
\def \nar {New Astron. Rev.} 

\def \JCAP {JCAP}
\def \araa {ARA\&A}
\def \aj {AJ}
\def \aar {A\&AR}
\def \apj {ApJ}
\def \apjl {ApJL}
\def \apjs {ApJS}
\def \asl {Adv. Sci. Lett.} 
\def \mnras {Mon.\ Non.\ Roy.\ Astron.\ Soc.}
\def \nat {Nat}
\def \pasj {PASJ}
\def \pasp {PASP}
\def \science {Science}

\def \gca {Geochim.\ Cosmochim.\ Acta}
\def \npa {Nucl.\ Phys.\ A}
\def \plb {Phys.\ Lett.\ B}
\def \prc {Phys.\ Rev.\ C}
\def \prd {Phys.\ Rev.\ D.}
\def \prl {Phys.\ Rev.\ Lett.}


\end{document}